\documentclass{article}
\usepackage{amsfonts}
\usepackage{amsmath}

\newtheorem{theorem}{Theorem}

\newtheorem{definition}[theorem]{Definition}

\newtheorem{proposition}[theorem]{Proposition}

\input{tcilatex}

\begin{document}

\title{On electromagnetism and generalized energy-momentum tensor of the
electromagnetic field in spaces with Finsler geometry}
\author{Nicoleta VOICU \\
Transilvania University, Brasov, Romania}
\maketitle

\begin{abstract}
By using variational calculus and exterior derivative formalism, we proposed
in \cite{animmath} and \cite{Cairo08} a new geometric approach to
electromagnetism in pseudo-Finsler spaces. In the present paper, we provide
more details, especially regarding generalized currents, the domain of
integration and gauge invariance. Also, for flat pseudo-Finsler spaces, we
define a generalized energy-momentum tensor consisting of two blocks, as the
symmetrized Noether current corresponding to the invariance of the field
Lagrangian with respect to spacetime translations. In curved spaces, one of
the blocks of the generalized energy-momentum tensor is obtained by varying
the field Lagrangian with respect to the metric tensor and the other one, by
varying the same Lagrangian with respect to the nonlinear connection.
\end{abstract}

\section{Introduction}

Classical electromagnetism is one of the most "rounded" theories of physics
and one has strong reasons to say that there is little to be added to it.
There exist several beautiful geometrical descriptions of this theory in
classical general relativity.

Still, what if spacetime is described not by Riemannian, but by Finslerian
geometry?\ As G. Yu Bogoslovski and H. Goenner noticed, "spacetime may be
not only in a state which is described by Riemann geometry but also in
states which are described by Finsler geometry". A lot of authors have
already considered Finslerian models for spacetime (see, for instance, \cite%
{Asanov}, \cite{Ba-St1}-\cite{Ba-St4}, \cite{Bogoslovski1}-\cite%
{Bogoslovski3}, \cite{Lagrange}, \cite{garas'ko}, \cite{Pavlov}, \cite{Rutz}%
, \cite{agd}, \cite{vacaru}, \cite{vacaru2}, \cite{Li}, in order to cite
just a few of them).

\bigskip

Regarding electromagnetism, we can expect that in spaces with Finslerian
geometry, the corresponding equations would change and that we might even
have to deal with some new quantities. In this paper, we will investigate
from a mathematical point of view these possible changes.

The first idea we must have in mind is that in spaces with Finsler geometry,
the metric tensor depends on the directional variables. Since Maxwell
equations involve the metric tensor, we notice that the solutions (hence,
the electromagnetic tensor) may also depend on these. This means that the
natural space to work on is not the spacetime manifold $M,$ but its tangent
bundle $TM.$

Thus, we extend the classical ideas of electromagnetic field theory to
Finsler spaces, as an application of the geometry of the tangent bundle $TM$.

Though we will speak throughout the chapter about Finsler spaces, all the
theory remains valid, with minimal changes, for more general anisotropic
spaces (Lagrange, generalized Lagrange ones).

This paper is a continuation of \cite{Cairo08} and \cite{animmath}. It is
based on variational calculus and classical methods in theoretical physics
(adapted to the tangent bundle). This approach offers an alternative to the
existing one by R. Miron and collaborators, \cite{Miron-Buchner}, \cite%
{Miron-Rad}, \cite{Lagrange}.

\section{A brief overview of the Riemannian case}

There are multiple definitions of the electromagnetic tensor, points of view
and formulations of the basic equations of electromagnetism on Riemannian
manifolds. Namely, the electromagnetic tensor can be regarded as the
curvature of a line bundle over the given manifold or it can be described in
terms of nonlinear/linear on $TM$ as in \cite{Miron-Rad}, or in terms of
differential forms.

In the following, we will adopt the language of differential forms, as it is
the most tightly related to variational calculus.

Let us consider a pseudo-Riemannian manifold $(M,g)$ of dimension 4, thought
of as spacetime manifold. We denote local coordinates on $M$ by $%
x=(x^{i})_{i=\overline{0,3}}$ and use the numbering from 0 to 3. The first
coordinate is regarded as the time coordinate and $(x^{\alpha })_{\alpha =%
\overline{1,3}},$ as spatial coordinates. As required by general relativity,
the metric $g=g(x)$ is supposed to have Lorentz signature $(+,-,-,-).$ Here
are some other notations and conventions we will use in the following:

- \thinspace Latin indices $i,j,k,...$ take values from 0 to 3; Greek
indices $\alpha ,\beta ,\gamma ,...$ take values from 1 to 3;

- for a vector field $v=(v^{i})_{i=\overline{0,3}}$ on $M,$ $\mathbf{v}$
will denote the spatial vector $\mathbf{v}=(v^{\alpha })_{\alpha =\overline{%
1,3}}.$

- $_{,k}$ -- partial derivative with respect to $\dfrac{\partial }{\partial
x^{k}};$

- $_{|k}$ -- Levi-Civita covariant derivative with respect to $\dfrac{%
\partial }{\partial x^{k}};$ $\gamma _{~jk}^{i}$ - Christoffel symbols of $%
g; $

- $g=\det (g_{ij});$ whenever it is not clear from the context whether we
refer to $g$ as the metric tensor or to the determinant of the corresponding
matrix, we will specify this;

- $\flat :TM\rightarrow T^{\ast }M,$ $\sharp :T^{\ast }M\rightarrow TM$ \ -
musical isomorphisms (lowering/raising indices);

- $d$ - exterior derivative of differential forms, $\ast $ - Hodge dual;

- $\epsilon _{i_{1}...._{i_{p}}}$ - signature of the permutation $%
(i_{1},....,i_{p});$

- $\nabla $ -- gradient taken with respect to the spatial coordinates $%
(x^{\alpha })$;

- $d^{4}x=dx^{0}\wedge dx^{1}\wedge dx^{2}\wedge dx^{3},$ $%
~~d^{3}x=dx^{1}\wedge dx^{2}\wedge dx^{3};$

- $d\Omega =\sqrt{|g|}d^{4}x,$ $dV=\dfrac{\sqrt{|g|}}{\sqrt{g_{00}}}d^{3}x$
- the invariant Riemannian volume element on spacetime and on the spatial
manifold respectively.

\bigskip

\subsection{Distances, volumes, divergence, codifferential}

Let us remind for the beginning some very quick facts about computation of
time intervals, distances and spatial volumes in general relativity (\cite%
{Landau}, pp. 315-320).

The (squared) arclength element $ds^{2}$ on the spacetime manifold $M$ can
be written as%
\begin{equation*}
ds^{2}=g_{00}(dx^{0})^{2}+2g_{0\alpha }dx^{0}dx^{\alpha }+g_{\alpha \beta
}dx^{\alpha }dx^{\beta }.
\end{equation*}

The spatial arclength element is defined as%
\begin{equation*}
dl^{2}=\gamma _{\alpha \beta }dx^{\alpha }dx^{\beta },~\ \ \ \ \ \ \ \gamma
_{\alpha \beta }=-g_{\alpha \beta }+\dfrac{g_{0\alpha }g_{0\beta }}{g_{00}}%
,~\ \ \ \alpha ,\beta =\overline{1,3}.
\end{equation*}

\bigskip

The determinant of the spacetime metric $g$ is%
\begin{equation*}
g=-g_{00}\gamma ;
\end{equation*}
we only consider reference frames for which both the determinant of the
spatial metric $\gamma $ and $g_{00}$ are positive:

$\gamma :=\det (\gamma _{ij})>0$ and $g_{00}>0.$

\bigskip

Here are some other relations we will use in the following.

The partial derivatives of $\sqrt{\left\vert g\right\vert }$ are given by:

\begin{equation*}
\dfrac{\partial (\ln \sqrt{\left\vert g\right\vert })}{\partial x^{j}}=%
\dfrac{1}{2}g^{ih}g_{ih,j}=\gamma _{~ji}^{i}.
\end{equation*}%
Consequently, the divergence{\footnotesize \ }$div(V)=\dfrac{1}{\sqrt{%
\left\vert g\right\vert }}\dfrac{\partial }{\partial x^{i}}(V^{i}\sqrt{%
\left\vert g\right\vert })$\ of a vector field can be written in terms of
covariant derivatives, as%
\begin{equation*}
div(V)=V_{~|i}^{i}.
\end{equation*}

Also, by expressing $g_{ij}$ as $\dfrac{1}{g^{-1}}\dfrac{\partial g^{-1}}{%
\partial g^{ij}},$ we get 
\begin{equation}
d(\ln \sqrt{\left\vert g\right\vert })=-\dfrac{1}{2}g_{ij}dg^{ij}.
\end{equation}

The latter equality is particularly useful when varying Lagrangians with
respect to the metric.

\bigskip

The codifferential of $\delta \xi =(-1)^{p}\ast ^{-1}d\ast \ $of a $p$-form $%
\xi =\dfrac{1}{p!}\xi _{i_{1}i_{2}...i_{p}}dx^{i_{1}}\wedge ...\wedge
dx^{i_{p}}$ is a $(p-1)$-form, with the property $\left\langle \eta ,\delta
\xi \right\rangle =\left\langle d\eta ,\xi \right\rangle ,$ where $%
\left\langle ~,~\right\rangle $ denotes the inner product of $p$-forms%
\footnote{%
The inner product of two $p$-forms $\theta =\theta
_{i_{1}...i_{p}}e^{i_{1}}\wedge ...\wedge e^{i_{p}}$ and $\psi =\psi
_{j_{1}...j_{p}}e^{j_{1}}\wedge ...\wedge e^{j_{p}}$ is given by $\int
g^{i_{1}j_{1}}...g^{i_{p}j_{p}}\theta _{i_{1}...i_{p}}\psi
_{j_{1}...j_{p}}d\Omega ,$ where the integral is taken on the whole manifold
(assuming that the integrands have compact support).}. For a 2-form, it is
given by%
\begin{equation*}
(\delta \xi )^{i}=\xi _{~~|j}^{ij}.
\end{equation*}

\subsection{4-potential and electromagnetic tensor}

The 4-potential is geometrically described in classical general relativity
as a 1-form 
\begin{equation}
A=A_{i}(x)dx^{i}.  \label{4-potential_classical}
\end{equation}

The electromagnetic tensor (or \textit{Faraday 2-form}) is described as the
2-form 
\begin{equation}
F=dA.  \label{em_tensor_classical}
\end{equation}

In local coordinates, this is 
\begin{equation}
F=\dfrac{1}{2}F_{jk}dx^{j}\wedge dx^{k},  \label{local_classical_em_tensor}
\end{equation}%
where%
\begin{equation}
F_{jk}=A_{k|j}-A_{j|k}.  \label{local_classical_em_tensor1}
\end{equation}

Due to the symmetry of the Levi-Civita connection, the latter expression can
be actually written in terms of partial derivatives only:%
\begin{equation}
F_{jk}=A_{k,j}-A_{j,k}.  \label{local_classical_em_tensor2}
\end{equation}

In the language of differential forms, the homogeneous Maxwell equations 
\begin{equation}
F_{ij|k}+F_{ki|j}+F_{jk|i}=0.  \label{hom_max_classical_covar}
\end{equation}

become%
\begin{equation}
dF=0.  \label{hom_max_classical}
\end{equation}

\bigskip

\textbf{Remark. }There exist two possible approaches regarding the potential 
$A$ and the electromagnetic tensor $F.$

\begin{enumerate}
\item One can consider as a fundamental object the electromagnetic tensor $%
F, $ regarded as a closed 2-form on the manifold. In this case, the
homogeneous Maxwell equation $dF=0$ , i.e., the closure condition for $F,$
is taken as an axiom. If the manifold $M$ is topologically "nice enough",
then one can apply Poincar\'{e}'s lemma, which entails the existence of a
1-form $A,$ such that $F=dA.$ That is, the existence of the 4-potential is
seen as a consequence of the homogeneous Maxwell equations.

\item Some authors consider the potential 1-form $A$ as a fundamental object
and \textit{define} $F$ as its exterior differential; in this approach, the
homogeneous Maxwell equation is obtained as an identity.
\end{enumerate}

Actually, for a Lagrangian theory of electromagnetism, it is essential to
have both a 1-form $A$ and a 2-form $F,$ related by (\ref{hom_max_classical}%
).

\bigskip

An important property of the electromagnetic field is \textit{gauge
invariance}. Namely, the field strength tensor $F$ is invariant to
transformations%
\begin{equation*}
A\mapsto A+d\psi ,
\end{equation*}%
where $\psi :M\rightarrow \mathbb{R}$ is a differentiable function.

\bigskip

\subsection{Lagrangian, equations of motion and inhomogeneous Maxwell
equations}

The second pair of Maxwell equations (inhomogeneous Maxwell equations) and
also, the equations of motion of charged particles in a given
electromagnetic field are obtained by variational methods.

The \textit{total action} attached to the field and to a system of particles
is%
\begin{equation}
S=-\sum mc\int ds-\sum \dfrac{q}{c}\int A_{k}(x)dx^{k}-\dfrac{1}{16\pi c}%
\int F_{ij}F^{ij}d\Omega .  \label{general_Lagrangian}
\end{equation}%
Here, $m,q,c$ are constants ($m$ denotes the mass of a particle, $q,$ its
charge, $c,$ the speed of light in vacuum), $d\Omega =\sqrt{|g|}d^{4}x$ is
the invariant volume element on spacetime and the sums are taken over the
particles in the system. The volume integral is taken over a bounded
interval of time and over the whole spatial manifold, under the assumption
that far away from sources, the field vanishes. Thus, we can actually think
the integral as taken over a "large enough" compact domain in $M$.

\bigskip

The first term, $S_{p}:=-\sum mc\int ds,$ corresponds to the Lagrangian $%
L_{p}:=-\sum mcds$ of the free particles.

The second term $S_{int}:=-\sum \dfrac{q}{c}\int A_{k}(x)dx^{k},$ given by
the Lagrangian $L_{int}:=-\sum \dfrac{q}{c}A_{k}(x)dx^{k}$ characterizes the
interaction between the particles and the field. The third term $S_{f}:=%
\dfrac{-1}{16\pi c}\int F_{ij}F^{ij}d\Omega =-\dfrac{1}{16\pi c}\int F\ast
F~d^{4}x$, characterizes the electromagnetic field in the given curved space.

By keeping the electromagnetic field fixed and varying the trajectory of a
particle in the action $S$ (which actually means varying the trajectory in $%
S_{p}+S_{int}$), one obtains the equations of motion of particles subject to
both gravitational and electromagnetic field, i.e., the expression of the
Lorentz force. If, conversely, in the action $S$ we keep trajectories fixed
and vary the electromagnetic field (which actually means to vary the
electromagnetic field in the sum $S_{int}+S_{f}$), we get the field
equations, i.e., the second pair of Maxwell equations.

\bigskip

Let us notice that $S_{p}$ and $S_{int}$ are line integrals, while $S_{f}$
is given by a volume integral. Hence, if we want to vary $S_{int}+S_{f},$ we
have to write this sum as a single volume integral, too. This is achieved by
means of the notion of charge density.

\textit{Charge density} $\rho $ is defined as the amount of electric charge
in a given spatial volume and it is basically a function of time and spatial
coordinates:%
\begin{equation*}
\rho =\rho (x).
\end{equation*}

The integral of $\rho $ over a certain region of space provides the total
charge situated inside that region:%
\begin{equation}
q=\int \rho dV,  \label{total_charge}
\end{equation}%
where $dV=\dfrac{\sqrt{\left\vert g\right\vert }}{\sqrt{g_{00}}}d^{3}x$ \ is
the spatial volume element. In this writing under an integral, it is
supposed that we actually see the charge distribution as "continuous". Total
charge is invariant to coordinate changes.

For a discrete distribution of charges $q_{1},...,q_{n}$ in a given volume,
we can still write the total charge in the form of the integral (\ref%
{total_charge}), if we define the charge density by means of the Dirac delta
function:%
\begin{equation*}
\rho =\underset{a=1}{\overset{n}{\sum }}\dfrac{q_{a}}{\sqrt{\gamma }}\delta (%
\mathbf{x-x}_{(a)}),
\end{equation*}%
where $\mathbf{x}=(x^{1},x^{2},x^{3})$ and $\mathbf{x}_{(a)}$ is the
position vector of the charge $q_{a}.$

\bigskip

By using relation (\ref{total_charge}), $S_{int}$ is written as%
\begin{equation*}
\mathit{\ }S_{int}=-\dfrac{1}{c}\int A_{i}J^{i}d\Omega ,
\end{equation*}%
where the quantities%
\begin{equation}
J^{i}:=\dfrac{\rho c}{\sqrt{g_{00}}}\dfrac{dx^{i}}{dx^{0}}
\label{current_classical}
\end{equation}%
are the components of a vector field $J$, called the \textit{4-current}.

Thus, the sum $S_{1}:=S_{int}+S_{f}$ can be written as a single integral as 
\begin{equation*}
S_{1}=-\int {\Large (}\dfrac{1}{c}A_{i}J^{i}+\dfrac{1}{16\pi c}F_{ij}F^{ij}%
{\Large )}\sqrt{\left\vert g\right\vert }d^{4}x.
\end{equation*}

\bigskip

By varying the above Lagrangian with respect to the potential $A,$ one gets
the field equations, i.e., the \textit{inhomogeneous Maxwell equation:}%
\begin{equation}
\delta F=-\dfrac{4\pi }{c}J_{\flat },  \label{inhom_max_classical}
\end{equation}%
or, in local writing,%
\begin{equation}
\dfrac{1}{\sqrt{\left\vert g\right\vert }}\dfrac{\partial }{\partial x^{j}}%
\left( F^{ij}\sqrt{\left\vert g\right\vert }\right) =-\dfrac{4\pi }{c}%
J^{i}~\ \ \Longleftrightarrow ~\ \ \ F_{~~|j}^{ij}=-\dfrac{4\pi }{c}J^{i}.
\label{inhom_max_classical_coords}
\end{equation}

\textbf{Remark. }The Maxwell equations and eventual choices of $\psi $ in
the transformations $A\mapsto A+d\psi $ do not completely determine the
potential $A.$ Hence, to $A,$ one can still impose supplementary conditions
(gauges). The most common is the \textit{Lorenz gauge }$A_{~|i}^{i}=0$;
under this condition the inhomogeneous Maxwell equations become:%
\begin{equation}
-A_{~~~|j}^{i|j}+A^{j}R_{~j}^{i}=-\dfrac{4\pi }{c}J^{i},
\label{Maxwell-de Rham}
\end{equation}%
where $R_{~j}^{i}=g^{ih}R_{h~jk}^{~k}$ are the components of the Ricci
tensor.

\bigskip

By using inhomogeneous Maxwell equation, one obtains that the 4-current $J$
identically satisfies the \textit{continuity equation:}%
\begin{equation}
div(J)=d(\ast J_{\flat })=0.  \label{continuity_eq_classical}
\end{equation}%
\textit{i.e.,} 
\begin{equation}
J_{~|i}^{i}=0.  \label{continuity_classical_coords}
\end{equation}

From a physical point of view, the continuity equation is equivalent to the 
\textit{charge conservation law}.

\bigskip

Let us now consider a single particle, subject to the action of a given
(fixed) electromagnetic field and determine the trajectory of this particle.
This can be achieved by varying the action $S$ with respect to the
trajectory. That is, we actually have to vary the action $%
S_{2}:=S_{p}+S_{int},$ written as a line intergral.

We\textbf{\ }notice that that the integral $S_{p}+S_{int}$ does not depend
on the choice of the parameter on the path of integration. Thus, we can
choose this parameter according to our wish. So, let us choose the arclength 
$s$ as a parameter, With this choice, we have $\left\Vert \dfrac{dx}{ds}%
\right\Vert =\sqrt{g_{ij}\dot{x}^{i}\dot{x}^{j}}=\dfrac{ds}{ds}=1.$

Thus, we can write

\begin{equation}
S_{2}=-mc\int ds-\dfrac{q}{c}\int A_{i}(x)dx^{i}=-\int (mc\sqrt{g_{ij}\dot{x}%
^{i}\dot{x}^{j}}+A_{i}\dot{x}^{i})ds,
\end{equation}%
where the integral is taken on some fixed compact interval $[s_{0},s_{1}].$

The Euler-Lagrange equations for the above Lagrangian are%
\begin{equation}
\dfrac{D\dot{x}^{i}}{ds}=\dfrac{q}{c}F_{~j}^{i}\dot{x}^{j},~\ \ i=\overline{%
0,3},  \label{Lorentz_eq_classical}
\end{equation}%
where $\dfrac{D\dot{x}^{i}}{ds}=\dfrac{d\dot{x}^{i}}{ds}+\gamma _{~jk}^{i}%
\dot{x}^{j}\dot{x}^{k}$ is the Levi-Civita covariant derivative.

The right hand sides of the above equations provide the expression of the 
\textit{Lorentz force} in the given curved space. Also, the first two terms
in (\ref{general_Lagrangian}) provide the canonical momentum%
\begin{equation}
p_{i}=mc\dfrac{\dot{x}_{i}}{\left\Vert \dot{x}\right\Vert }+\dfrac{q}{c}%
A_{i}.  \label{canonical_momentum}
\end{equation}

\subsection{Energy-momentum tensor}

Another important quantity in general relativity, is \textit{energy-momentum
tensor} (or \textit{stress-energy tensor}) $T.$ In classical general
relativity, the energy-momentum tensor is symmetric and, for a closed
system, its covariant divergence is zero.

\bigskip

\textbf{A. In flat Minkowski space}

In the Minkowski space $(\mathbb{R}^{4},\eta )$ ($\eta =diag(1,-1,-1,-1)$)
of special relativity, it makes sense to speak about spacetime translations%
\begin{equation*}
x\mapsto x+a~\ \ \ (a\text{ - constant 4-vector}).
\end{equation*}%
The Lagrangians $L_{f},L_{int},L_{p}$ in (\ref{general_Lagrangian}) are all
invariant with respect to these translations (this can be easily checked, by
noticing that neither of them depends explicitely on the spacetime
coordinates $x^{i}$).

Moreover, in this case we have $g=\det (\eta _{ij})=-1$ and covariant
derivatives coincide with partial ones.

\bigskip

According to Noether's theorem, the invariance of an action 
\begin{equation*}
S=\dfrac{1}{c}\int \Lambda (q_{(l)},\dfrac{\partial q_{(l)}}{\partial x^{i}}%
)d\Omega ,
\end{equation*}
to translations implies that the quantities%
\begin{equation*}
\tilde{T}_{~i}^{k}=q_{(l),i}\dfrac{\partial \Lambda }{\partial q_{(l),k}}%
-\delta _{i}^{k}\Lambda =0
\end{equation*}%
are conserved ($div\tilde{T}=0$). They define a tensor of rank two (the 
\textit{Noether current} attached to the Lagrangian).

The Noether current is generally not symmetric. Still, this situation can be
"mended"\ by adding a divergence term: 
\begin{equation*}
T^{ik}=\tilde{T}^{ik}+\dfrac{\partial \psi ^{ikl}}{\partial x^{l}},
\end{equation*}%
(where $\psi ^{ikl}(x)=-\psi ^{ilk}(x)$ are functions of class at least
two), which does not affect the value of the action integral (assuming, as
usually, that on the boundary of the integration domain, the involved
functions vanish). Thus, one obtains a symmetric tensor $T$ of rank two,
with 
\begin{equation*}
\dfrac{\partial T_{~i}^{k}}{\partial x^{k}}=0
\end{equation*}

The \textit{energy-momentum tensor of the electromagnetic field }in flat
space is defined as the symmetrized Noether current given by the invariance
of the field Lagrangian $L_{f}$ to spacetime translations.

For electromagnetism, we have $q_{(k)}=A_{(k)}$ and 
\begin{equation*}
\Lambda =-\dfrac{1}{16\pi }F_{ij}F^{ij}.
\end{equation*}

By supposing, at first, that $\rho =0$ (which implies $J=0$), we have:%
\textbf{\ }$\tilde{T}_{~i}^{l}=\dfrac{1}{4\pi }(-F^{lk}A_{k,i}+\dfrac{1}{4}%
\delta _{i}^{l}F_{jk}F^{jk});$ by adding the quantity $\dfrac{1}{4\pi }%
A_{~,l}^{i}F^{kl}=\dfrac{1}{4\pi }(A^{i}F^{kl})_{,l}$), one gets the
energy-momentum tensor as: 
\begin{equation}
T_{~i}^{l}=\dfrac{1}{4\pi }(-F^{lk}F_{ik}+\dfrac{1}{4}\delta
_{i}^{l}F_{jk}F^{jk}).  \label{sem_tensor_classical}
\end{equation}

Thus, if $J=0$, then 
\begin{equation*}
div(T)=0.
\end{equation*}

In the situation when we have charged matter ($J\not=0$), the
energy-momentum tensor satisfies the identities:%
\begin{equation}
T_{~i,j}^{j}=-\dfrac{1}{c}F_{ij}J^{j}
\end{equation}%
(which can be proved by means of Maxwell equations). In brief, 
\begin{equation}
div(T)=-\dfrac{1}{c}i_{J}F.
\end{equation}

The quantity $\dfrac{1}{c}i_{J}F$ is called the \textit{density of Lorentz
force}.

\bigskip

\textbf{B. In curved spaces}

In general Relativity, the energy-momentum tensor $T$ is defined by the
relation%
\begin{equation}
\delta _{g}S=\dfrac{1}{2c}\int T_{ik}\delta g^{ik}d\Omega =-\dfrac{1}{2c}%
\int T^{ik}\delta g_{ik}d\Omega .  \label{def_sem_tensor_riem}
\end{equation}

Variation with respect to the metric leads to: 
\begin{equation}
T_{ij}=\dfrac{1}{4\pi }(-F_{j}^{~k}F_{ik}+\dfrac{1}{4}g_{ij}F_{lk}F^{lk}),
\label{sem_tensor_riem}
\end{equation}%
which agrees to the expression of the energy-momentum tensor in flat space.

\bigskip

Again, by using (both homogeneous and inhomogeneous) Maxwell equations, one
gets that in curved pseudo-Riemannian spaces, the covariant divergence of
the stress-energy tensor of the electromagnetic field is equal to minus the
density of Lorentz force: 
\begin{equation}
T_{~i|j}^{j}=-\dfrac{1}{c}F_{ij}J^{j}.
\end{equation}

\textbf{Conclusion. }In a geometric language, the above fundamental
equations of electromagnetic field theory can be written briefly as:

\begin{itemize}
\item \thinspace $F=dA.$

\item $dF=0$ -- homogeneous Maxwell equation;

\item $\delta F=-\dfrac{4\pi }{c}J_{\flat }$ -- inhomogeneous Maxwell
equation;

\item $div(J)=0$ -- continuity equation;

\item $\nabla \cdot T=-\dfrac{1}{c}i_{J}F$ -- energy-momentum conservation
(where $\nabla \cdot T$ denotes covariant divergence).
\end{itemize}

\bigskip

\section{Some geometric structures in Finsler spaces}

Let, again, $M$ be a 4-dimensional differentiable manifold of class $%
\mathcal{C}^{\infty },$ thought of as spacetime manifold. This time we will
also speak about the tangent bundle $(TM,\pi ,M)$ and denote $%
(x^{i},y^{i})_{i=\overline{0,3}}$ the coordinates in a local chart on $TM;$
we preserve the notations in the previous section, with the only difference
that instead of Levi-Civita covariant derivatives, we will use other
covariant derivation laws. Also, we denote partial derivation with respect
to $y^{i}$ by a dot: $_{\cdot i}.$ We will sometimes call the base
coordinates $x^{i}$ positional variables and the fiber ones, directional
variables.

\bigskip

A \textit{Finsler fundamental function} on $M,$ is a function $\mathcal{F}%
:TM\rightarrow \mathbb{R}$ with the properties, \cite{Szilasi}:

\begin{enumerate}
\item $\mathcal{F=F}(x,y)$ is smooth for $y\not=0;$

\item $\mathcal{F}$ is positive homogeneous of degree 1, i.e., $\mathcal{F}%
(x,\lambda y)=\lambda \mathcal{F}(x,y)$ for all $\lambda >0;$

\item The \textit{Finslerian metric tensor}:\textit{\ } 
\begin{equation}
g_{ij}(x,y)=\dfrac{1}{2}\dfrac{\partial ^{2}\mathcal{F}^{2}}{\partial
y^{i}\partial y^{j}},  \label{Finsler_metric}
\end{equation}%
is nondegenerate: $\det (g_{ij}(x,y))\not=0,~\forall x\in M,$ $y\in
T_{x}M\backslash \{0\}.$
\end{enumerate}

In the following, we will consider that the metric has signature $(+,-,-,-).$

\bigskip

\textbf{Remark. }Strictly speaking, it would be more rigorous to preserve
the term "Finslerian" for the case when the metric tensor $g$ is positive
definite and to call "pseudo-Finsler" the spaces in which the metric is only
nondegenerate (of constant signature). But since a lot of authors already
use in the latter case the term \textit{Finsler}, we will also adopt this
more relaxed terminology.

\bigskip

In a Finsler space, the element of arc length along a curve $t\mapsto x(t)$
is%
\begin{equation*}
ds=\mathcal{F}(x,\dfrac{dx}{dt})dt.
\end{equation*}

Finsler spaces are a generalization of pseudo-Riemannian manifolds, in which
the coefficients of the metric tensor are no longer functions defined on $M$%
, but on the tangent bundle $TM.$ Actually, if on a pseudo-Riemannian
manifold, the tangent space at each point carries a pseudo-Euclidean metric
structure, in a Finsler space, at each fixed point $x^{0},$ the "norm" $%
\mathcal{F}(x_{0},y)$ is generally, not given by a quadratic form.
1-homogeneity of $\mathcal{F}$ in $y$ insures that the integral $\int ds$
does not depend on eventual changes of the parameter along the curve (hence,
the notion of arclength $s$ is uniquely defined, no matter from which
initial parametrer $t$ we start).

\bigskip

Given a (pseudo-)Finslerian metric tensor $g_{ij}=g_{ij}(x,y),$ the
corresponding spatial metric is defined similarly to the pseudo-Riemannian
case: $\gamma _{\alpha \beta }=-g_{\alpha \beta }+\dfrac{g_{0\alpha
}g_{0\beta }}{g_{00}},$ $\alpha ,\beta \in \{1,2,3\}$ and its determinant is 
$\det (\gamma _{\alpha \beta })=\dfrac{\sqrt{|g|}}{\sqrt{g_{00}}}.$

With respect to coordinate changes on the tangent bundle $TM$ induced by
coordinate changes $(x^{i})\mapsto (\tilde{x}^{i})$ on the base manifold $M,$
i.e., under coordinate changes%
\begin{equation}
\tilde{x}^{i}=\tilde{x}^{i}(x),~\ \ \ \tilde{y}^{i}=\dfrac{\partial \tilde{x}%
^{i}}{\partial x^{j}}y^{j}  \label{coord_change_TM}
\end{equation}%
the quantities $\dfrac{\partial }{\partial y^{i}}$ have a tensorial rule of
transformation: $\dfrac{\partial }{\partial y^{i}}=\dfrac{\partial \tilde{x}%
^{j}}{\partial x^{i}}\dfrac{\partial }{\partial \tilde{y}^{j}},$ while the
quantities $\dfrac{\partial }{\partial x^{i}}$ have a more complicated
transformation law, \cite{Shen}, \cite{Lagrange}.

If we want to work with tensorial\textit{\ }blocks only, then we have to use
Ehresmann (nonlinear) connections on $TM.$

\bigskip

We will denote by $(N_{~i}^{\bar{j}})$ the coefficients of an Ehresmann
connection, and by 
\begin{equation}
\begin{array}{c}
\delta _{i}=\dfrac{\partial }{\partial x^{i}}-N_{~i}^{\bar{\imath}}\dfrac{%
\partial }{\partial y^{\bar{\imath}}},~\ \ \partial _{\bar{\imath}}=\dfrac{%
\partial }{\partial y^{\bar{\imath}}}, \\ 
(dx^{i},\delta y^{\bar{\imath}}=dy^{\bar{\imath}}+N_{~j}^{\bar{\imath}%
}dx^{j})%
\end{array}%
\end{equation}%
the elements of the corresponding \textit{adapted basis }and of its dual
cobasis. Thus, with respect to coordinate changes (\ref{coord_change_TM}), $%
\delta _{i}$ and $\partial _{\bar{\imath}}$ have tensorial rules of
transformation, i.e., $\delta _{i}=\dfrac{\partial x^{j^{\prime }}}{\partial
x^{i}}\delta _{j^{\prime }},$ $\partial _{\bar{\imath}}=\dfrac{\partial
x^{j^{\prime }}}{\partial x^{\bar{\imath}}}\partial _{j^{\prime }}.$

\bigskip

In the adapted basis, any vector field $V$ on $TM$ can be written as $%
V=V^{i}\delta _{i}+V^{\bar{\imath}}\partial _{\bar{\imath}};$ the component%
\begin{equation*}
hV=V^{i}\delta _{i}
\end{equation*}%
is a vector field, called the \textit{horizontal }component of $V,$ while%
\begin{equation*}
vV=V^{\bar{\imath}}\partial _{\bar{\imath}}
\end{equation*}%
is also a vector field, called its \textit{vertical }component. Similarly, a
1-form $\omega $ on $TM$ can be decomposed into invariant blocks as $\omega
=\omega _{i}dx^{i}+\omega _{\bar{\imath}}\delta y^{\bar{\imath}},$ with%
\begin{equation*}
h\omega =\omega _{i}dx^{i}
\end{equation*}%
called the \textit{horizontal} component, and%
\begin{equation*}
v\omega =\omega _{\bar{\imath}}\delta y^{\bar{\imath}}
\end{equation*}%
the \textit{vertical }one, \cite{Lagrange}\textit{. }Accordingly, any tensor
field on $TM$ is decomposed with respect to the Ehresmann connection into
invariant blocks.

In the following, whenever needed to make a clear distinction, we will
denote by $i,j,k,...$ indices corresponding to horizontal geometric objects
and by $\bar{\imath},$ $\bar{j},$ $\bar{k},...$ (with bars), indices
corresponding to vertical ones - though, unless needed, we will not be too
strict in this respect. By capital letters $A,B,C,...$ we will always denote
indices which take values corresponding to both distributions: $A,B,C,...\in
\{i,j,k,...,\bar{\imath},$ $\bar{j},$ $\bar{k},...\}$

\bigskip

Let us complete $g$ up to a block metric (an \textit{hv-metric, }\cite%
{Lagrange}) on $TM:$%
\begin{equation}
G_{AB}(x,y)=g_{ij}(x,y)dx^{i}\otimes dx^{j}+v_{\bar{\imath}\bar{j}%
}(x,y)\delta y^{\bar{\imath}}\otimes \delta y^{\bar{j}}.  \label{metric}
\end{equation}%
where $g$ is the given Finslerian metric tensor and $v$ is a positive
definite metric tensor\footnote{%
Assuming that the topological space $M$ is metrizable, it appears as
advantageous to choose, for instance, a metric $v$ which provides the
topology of $M.$ In the case when $(M,g)$ is the Minkowski space, the
manifold topology of $M=\mathbb{R}^{4}$ is the Euclidean one, hence we can
choose $v$ as the Euclidean metric.}. Thus, $(TM,G)$ becomes a
pseudo-Riemannian space and we can speak about the Riemannian (invariant)
volume element on $TM:$%
\begin{equation*}
d\Omega =\sqrt{\left\vert G\right\vert }d^{4}x\wedge d^{4}y.
\end{equation*}%
where $G=\det (G_{AB})$ (we have written $d^{4}y$ instead of $\delta ^{4}y$
in the above exterior product, since $d^{4}x\wedge d^{4}y=d^{4}x\wedge
\delta ^{4}y$). The determinant $G$ is, obviously, 
\begin{equation*}
G=g\cdot v,~\ \ g=\det (g_{ij}),~\ v=\det (v_{\bar{\imath}\bar{j}}).
\end{equation*}

The volume element $d\Omega $ defines a volume element $d\Omega _{M}$ on $M$
by:%
\begin{equation*}
d\Omega _{M}=\sigma (x)d^{4}x,~~\ \ \ \ \sigma (x)=\underset{D_{x}}{\int }%
\sqrt{\left\vert G\right\vert }d^{4}x\wedge d^{4}y,
\end{equation*}%
where $D_{x}=\{y\in T_{x}M~|~v_{ij}(x,y)y^{i}y^{j}\leq r^{2}\}$ and $r=\sqrt[%
4]{2/\pi ^{2}}$ is chosen such that the 3-sphere of radius $r$ in the
4-dimensional Euclidean space has the volume equal to 1. This volume element
generalizes the idea of Holmes-Thompson volume in \cite{Shen1}\footnote{%
The classical idea of Holmes-Thompson volume involves integration on the
indicatrix $I_{x}=\{y\in T_{x}M|~g_{ij}y^{i}y^{j}=1\}.$ If the Finsler
metric $g$ is not positive definite, the indicatrix $I_{x}$ is generally
non-compact, hence this classical idea cannot be applied in our case.
Choosing as vertical part $v$ of the metric $G$ on $TM$ a positive definite
one (for instance, related to the spacetime topology) and integrating on
balls given by the metric $v$ solves this problem.}. Regarding integration
with respect to $x$, we can assume that the corresponding domain is a "large
enough" compact one (in the assumption that far away from sources, the field
is negligible and the considered time interval is a bounded one).

Having a metric structure on $TM$, there now make sense notions such as:
Hodge dual $\ast $ or codifferential $\delta $ of $p$-forms on $TM,$
gradient of a function and divergence of a vector field.

\bigskip

The \textit{divergence} of a vector field $V=V^{i}\delta _{i}+V^{\bar{\imath}%
}\partial _{\bar{\imath}}$ on $TM$ is obtained from the relation $\mathcal{L}%
_{V}d\Omega =divVd\Omega $. In the adapted frame to an arbitrary nonlinear
conection, the divergence of a vector field is expressed as%
\begin{equation}
divV=\dfrac{1}{\sqrt{\left\vert G\right\vert }}\left[ \delta _{i}(V^{i}\sqrt{%
|G|})+\partial _{\bar{\imath}}(V^{\bar{\imath}}\sqrt{\left\vert G\right\vert 
})\right] -N_{~i\cdot \bar{j}}^{\bar{j}}V^{i}.  \label{divergence}
\end{equation}

\bigskip

In particular, if the vertical block of the metric is a Riemannian one $%
v=v(x),$ then:

- the functions%
\begin{equation}
\overset{(v)}{N}\overset{}{_{~\bar{j}}^{\bar{\imath}}}=\overset{(v)}{\gamma }%
\overset{}{_{~jk}^{\bar{\imath}}(x)}y^{k},  \label{special_connection}
\end{equation}%
where $\overset{(v)}{\gamma }\overset{}{_{~jk}^{i}(x)}$ are the Christoffel
symbols of $v,$ are the coefficients of a nonlinear connection on $M;$

- in terms of this nonlinear connection, the expression of the divergence is
simplified as:%
\begin{equation}
divV=\dfrac{1}{\sqrt{\left\vert g\right\vert }}\left[ \delta _{i}(V^{i}\sqrt{%
|g|})+\partial _{\bar{\imath}}(V^{\bar{\imath}}\sqrt{\left\vert g\right\vert 
})\right] .
\end{equation}

\bigskip

The codifferential of any $p$-form $\xi =\dfrac{1}{p!}\xi
_{i_{1}i_{2}...i_{p}}e^{i_{1}}\wedge ...\wedge e^{i_{p}}$ on $TM$\ is the $%
(p-1)$-form $\delta \xi =(-1)^{p}\ast ^{-1}d\ast ;$ it\ can be also
calculated from the relation $\left\langle \eta ,\delta \xi \right\rangle
=\left\langle d\eta ,\xi \right\rangle ,$ where $\left\langle
~,~\right\rangle $ denotes the inner product of $p$-forms\footnote{%
The inner product of two $p$-forms $\theta =\theta
_{i_{1}...i_{p}}e^{i_{1}}\wedge ...\wedge e^{i_{p}}$ and $\psi =\psi
_{j_{1}...j_{p}}e^{j_{1}}\wedge ...\wedge e^{j_{p}}$ is traditionally given
by $\int g^{i_{1}j_{1}}...g^{i_{p}j_{p}}\theta _{i_{1}...i_{p}}\psi
_{j_{1}...j_{p}}d\Omega ,$ where the integral is taken on the whole manifold
(and it makes sense, for instance, for objects with compact support). In the
case of $TM,$ we will not integrate on the whole $TM,$ but on a compact
domain as specified above.}.

For a 2-form%
\begin{equation*}
\xi =\dfrac{1}{2}\xi _{ij}dx^{i}\wedge dx^{j}+\xi _{ia}dx^{i}\wedge \delta
y^{a}+\dfrac{1}{2}\xi _{ab}\delta y^{a}\wedge \delta y^{b}
\end{equation*}%
on $TM,$ the codifferential of $\xi $ is a 1-form $\delta \xi =\omega
_{i}dx^{i}+\omega _{a}\delta y^{a}$ whose contravariant components are given
by:%
\begin{eqnarray*}
\omega ^{i} &=&\dfrac{1}{\sqrt{\left\vert G\right\vert }}{\Large [}\delta
_{j}(\xi ^{ij}\sqrt{\left\vert G\right\vert })+\partial _{\bar{j}}(\xi ^{i%
\bar{j}}\sqrt{\left\vert G\right\vert }){\Large ]}-\xi ^{ij}N_{~j\cdot \bar{k%
}}^{\bar{k}}; \\
\omega ^{\bar{\imath}} &=&\dfrac{1}{\sqrt{\left\vert G\right\vert }}{\Large [%
}\delta _{j}(\xi ^{\bar{\imath}j}\sqrt{\left\vert G\right\vert })+\partial _{%
\bar{j}}(\xi ^{\bar{\imath}\bar{j}}\sqrt{\left\vert G\right\vert }){\Large ]}%
-\dfrac{1}{2}\xi ^{jk}R_{~jk}^{\bar{\imath}}-\xi ^{\bar{\imath}j}N_{~j\cdot 
\bar{k}}^{\bar{k}}+\xi ^{\bar{k}j}N_{~j\cdot \bar{k}}^{\bar{\imath}}.
\end{eqnarray*}

\bigskip

Choosing a nonlinear connection and a notion of covariant derivation or
another can help to express locally in a more or less elegant form the
obtained equations.

It appears as convenient to choose the following linear connection $D\Gamma
(N)$, inspired from \cite{Lagrange} (just - with a different covariant
derivation law for vertical fields)\footnote{%
The choice of this linear connection instead of the classical metrical
linear connection, \cite{Lagrange}, appeared as a little bit more
comfortable when expressing, for instance, the homogeneous Maxwell equation
in coordinates. This is just an example. All the results can be re-expressed
in terms of other linear connections.}:%
\begin{eqnarray}
X_{~|i}^{j} &=&\delta _{i}X^{j}+L_{~hi}^{j}X^{h},~~\ X_{~\cdot \bar{\imath}%
}^{j}=\dfrac{\partial X^{j}}{\partial y^{\bar{\imath}}},
\label{covariant_derivs} \\
X_{~|i}^{\bar{j}} &=&\delta _{i}X^{\bar{j}}+L\overset{}{_{~\bar{h}i}^{\bar{j}%
}}X^{\bar{h}},~~\ X_{~\cdot \bar{\imath}}^{\bar{j}}=\dfrac{\partial X^{\bar{j%
}}}{\partial y^{\bar{\imath}}},  \notag
\end{eqnarray}%
where 
\begin{eqnarray}
L_{~jk}^{i} &=&\dfrac{1}{2}g^{ih}(\delta _{k}g_{hj}+\delta _{j}g_{hk}-\delta
_{h}g_{jk}),  \label{Chern} \\
L\overset{}{_{~\bar{j}k}^{\bar{\imath}}} &=&N_{~k\cdot \bar{j}}^{\bar{\imath}%
}+\dfrac{1}{2}v^{\bar{\imath}\bar{h}}(\delta _{k}v_{\bar{h}\bar{j}%
}-N_{~k\cdot \bar{j}}^{\bar{l}}v_{\bar{l}\bar{h}}-N_{~k\cdot \bar{h}}^{\bar{l%
}}v_{\bar{l}\bar{j}}).  \notag
\end{eqnarray}

The above linear connection is a \textit{distinguished connection}, \cite%
{Lagrange}, meaning that it preserves the distributions generated by the
Ehresmann connection $N$ and it is \textit{h}-metrical, i.e., $g_{ij|k}=0,$ $%
v_{\bar{\imath}\bar{j}|k}=0,$ $\forall i,j,k,\bar{\imath},\bar{j}\in \{0,3\}.
$ Its only nonvanishing components of its torsion tensor $T$ are%
\begin{eqnarray*}
R_{~jk}^{\bar{\imath}} &=&\delta y^{\bar{\imath}}(T(\delta _{k},\delta
_{j}))=\delta _{k}N_{~j}^{\bar{\imath}}-\delta _{j}N_{~k}^{\bar{\imath}}; \\
P_{~j\bar{k}}^{\bar{\imath}} &=&\delta y^{\bar{\imath}}(T(\partial _{\bar{k}%
},\delta _{j}))=N_{~j\cdot \bar{k}}^{\bar{\imath}}-L_{~\bar{k}j}^{\bar{\imath%
}}.
\end{eqnarray*}

\bigskip

For the linear connection $D\Gamma (N)$ above defined, there hold the
relations:

\begin{equation}
\delta _{j}(\ln \sqrt{\left\vert g\right\vert })=L_{~ji}^{i},~\ \delta
_{j}(\ln \sqrt{\left\vert v\right\vert })-N_{~j\cdot \bar{k}}^{\bar{\imath}%
}=-P_{~i\bar{j}}^{\bar{\imath}}.  \label{derivs_det}
\end{equation}

Consequently, the divergence of a horizontal vector field $V^{H}=V^{i}\delta
_{i}$ on $TM$ can be written as:%
\begin{equation}
div(V^{H})=(V_{~|i}^{i}-P_{~i\bar{j}}^{\bar{j}}V^{i}).
\label{diverg_horizontal}
\end{equation}

Another important notion for a Finsler space is the \textit{Cartan tensor }$%
C $ given by%
\begin{equation*}
C_{~j\bar{k}}^{i}=\dfrac{1}{2}g^{ih}g_{hj\cdot \bar{k}}.
\end{equation*}%
It also has the property that%
\begin{equation*}
\dfrac{\partial (\ln \sqrt{\left\vert g\right\vert })}{\partial y^{\bar{j}}}%
=C_{~i\bar{j}}^{i}.
\end{equation*}

\bigskip

\textbf{Particular case: }If $v$ is a Riemannian metric, i.e., if%
\begin{equation}
G_{AB}(x,y)=g_{ij}(x,y)dx^{i}\otimes dx^{j}+v_{\bar{\imath}\bar{j}}(x)\delta
y^{\bar{\imath}}\otimes \delta y^{\bar{j}}.
\end{equation}%
and $\overset{v}{N}$ is given by (\ref{special_connection}), then%
\begin{equation*}
L_{~\bar{j}k}^{\bar{\imath}}=\overset{(v)}{\gamma }\overset{}{_{~jk}^{i}}%
=N_{~k\cdot \bar{j}}^{\bar{\imath}}
\end{equation*}%
and the only nonvanishing components of the torsion remain $R_{~jk}^{\bar{%
\imath}}=r_{l~jk}^{~i}(x)y^{l},$ where $r_{l~jk}^{~i}$ are the local
components of the curvature of the Levi-Civita connection of $v.$ Moreover,
in this case,%
\begin{equation*}
\delta _{j}(\ln \sqrt{\left\vert v\right\vert })=\overset{(v)}{\gamma }%
\overset{}{_{~ji}^{i}},~P_{~i\bar{j}}^{\bar{\imath}}=0.
\end{equation*}%
This will simplify a lot of calculations. For instance\textbf{,} the
divergence of a vector field $V=V^{i}\delta _{i}+V^{\bar{\imath}}\partial _{%
\bar{\imath}}$ and the codifferential $\omega =\delta \xi $ of a 2-form $\xi
=\dfrac{1}{2}\xi _{ij}dx^{i}\wedge dx^{j}+\xi _{i\bar{j}}dx^{i}\wedge \delta
y^{\bar{j}}+\dfrac{1}{2}\xi _{\bar{\imath}\bar{j}}\delta y^{\bar{\imath}%
}\wedge \delta y^{\bar{j}}$ on $TM$ are given in terms of covariant
derivatives (\ref{Chern}):%
\begin{equation*}
divV=V_{~|i}^{i}+V_{~\cdot \bar{\imath}}^{\bar{\imath}}+V^{\bar{\imath}}C_{~j%
\bar{\imath}}^{j},
\end{equation*}

and 
\begin{eqnarray*}
\omega ^{i} &=&\xi _{~~|j}^{ij}+\xi _{~~\cdot \bar{j}}^{i\bar{j}}+\xi ^{i%
\bar{j}}C_{~l\bar{j}}^{l}; \\
\omega ^{\bar{\imath}} &=&\xi _{~~|j}^{\bar{\imath}j}+\xi _{~~\cdot \bar{j}%
}^{\bar{\imath}\bar{j}}+\xi ^{\bar{\imath}\bar{j}}C_{~l\bar{j}}^{l}-\dfrac{1%
}{2}\xi ^{jk}R_{~jk}^{\bar{\imath}}.
\end{eqnarray*}

\section{4-potential 1-form}

If we want to use variational calculus in order to provide a generalization
of electromagnetic field theory to Finsler spaces, we need a generalization
of the notion of 4-potential.

Let us now see how does the notion of 4-potential transform in the case when
the geometry of the space is no longer Riemannian, but Finslerian, i.e., when%
\begin{equation*}
g_{ij}=g_{ij}(x,y).
\end{equation*}

We notice that the inhomogeneous Maxwell equations involve the components of
the metric tensor, which depend on the fiber coordinates $y^{i}.$It becomes
clear that generally, the solutions $A$ would depend on both $x$ and $y.$
Also, the equations themselves could become more complicated.

\bigskip

Consequently, from now on, we will consider 
\begin{equation}
A=A(x,y).  \label{anisotropic_potential}
\end{equation}

For reasons which will be clarified later, we will also assume that the
components $A_{i}$ are 0-homogeneous in $y:$%
\begin{equation*}
A_{i}(x,\lambda y)=A_{i}(x,y).
\end{equation*}%
That is, we will allow $A$ to depend on the direction of $y$, but not on its
magnitude.

In the following, we will focus on the action (\ref{general_Lagrangian}) and
determine the consequences of the $y$-dependence of the metric $g$ - and of
the potential $A.$ In pseudo-Finslerian spaces, the first term $S_{p}$
formally remains the same, with the only difference that in the expression $%
ds^{2}=g_{ij}(x,y)y^{i}y^{j}dt,$ $g_{ij}$ depends on $y=\dot{x}.$

\bigskip

\section{Faraday 2-form and homogeneous Maxwell equations}

Let us define the generalized Faraday 2-form (the electromagnetic tensor) in
the same way as in Riemannian spaces: 
\begin{equation}
F=dA;  \label{F-A}
\end{equation}%
in local coordinates, this is%
\begin{equation}
F:=\dfrac{1}{2}F_{ij}dx^{i}\wedge dx^{j}+F_{i\bar{j}}dx^{i}\wedge \delta y^{%
\bar{j}},  \label{def_F}
\end{equation}

In terms of adapted derivatives, components of the Faraday 2-form are
expressed as%
\begin{equation*}
F_{ij}=\delta _{i}A_{j}-\delta _{j}A_{i},~\ \ F_{i\bar{j}}=-~\partial _{\bar{%
j}}A_{i}
\end{equation*}%
and in terms of covariant derivatives (\ref{covariant_derivs}), we get 
\begin{equation}
F_{ij}=A_{j|i}-A_{i|j},~\ \ F_{i\bar{j}}=-A_{i\cdot \bar{j}}.  \label{F}
\end{equation}

\bigskip

In particular, if $A=A(x)$ does not depend on the directional variables, we
get $F=\dfrac{1}{2}(A_{j|i}-A_{i|j})dx^{i}\wedge dx^{j},$ which is similar
to the expression in \cite{Lagrange}, \cite{Miron-Rad}.

\bigskip

The electromagnetic tensor $F$ remains invariant under transformations 
\begin{equation}
A(x,y)~\mapsto A(x,y)+d\lambda (x),  \label{gauge_invar}
\end{equation}%
where $\lambda :M\rightarrow \mathbb{R}$ is a scalar function, since $%
d(A+d\lambda )=dA+d(d\lambda )=dA.$

\bigskip

Since $F$ is, by definition, a closed 2-form, its exterior derivative
identically vanishes. In other words:

\begin{proposition}
There holds the generalized homogeneous Maxwell equation: 
\begin{equation}
dF=0.  \label{Max1}
\end{equation}
\end{proposition}

Obviously, in local coordinates, equation (\ref{Max1}) will acquire
different forms, depending on the chosen Ehresmann connection $N$ and linear
connection $D\Gamma (N)$.

\bigskip

In terms of covariant derivatives (\ref{Chern}), equation (\ref{Max1}) is
read as: 
\begin{eqnarray*}
&&F_{ij|k}+F_{ki|j}+F_{jk|i}=-\underset{(i,j,k)}{\sum }R_{~jk}^{\bar{h}}F_{i%
\bar{h}}; \\
&&F_{\bar{\imath}j|k}+F_{k\bar{\imath}|j}+F_{jk\cdot \bar{\imath}}=P_{~j\bar{%
\imath}}^{\bar{h}}F_{k\bar{h}}-P_{~k\bar{\imath}}^{\bar{h}}F_{j\bar{h}},~\ \
F_{k\bar{\imath}\cdot \bar{j}}+F_{\bar{j}k\cdot \bar{\imath}}=0.
\end{eqnarray*}

If $v=v(x)$ and $N_{~j}^{\bar{\imath}}(x,y)=\overset{(v)}{\gamma }\overset{}{%
_{~jk}^{\bar{\imath}}}y^{k}$, then the second set of equations becomes 
\begin{equation*}
F_{\bar{\imath}j|k}+F_{k\bar{\imath}|j}+F_{jk\cdot \bar{\imath}}=0.
\end{equation*}

The first set in the above is the analogue (in the nonholonomic frame $%
(\delta _{i},\partial _{\bar{\imath}})$ on $TM$) of the usual homogeneous
Maxwell equations. In the cases when we can choose the nonlinear connection $%
N$ such that the horizontal distribution is integrable, then also the right
hand sides of the first set of equations vanish.

\bigskip

In the above, we have started from $A$ as an \textit{a priori} given object
and defined $F$ as its exterior derivative. Let us now proceed conversely
and suppose that $F$ is given. As we have shown in (\cite{animmath}), under
the assumptions that: the manifold $M$ is contractible and~$F$ is a closed
2-form with vanishing $\delta y^{i}\wedge \delta y^{j}$ component, i.e., 
\begin{equation*}
F:=\dfrac{1}{2}F_{ij}dx^{i}\wedge dx^{j}+F_{i\bar{j}}dx^{i}\wedge \delta y^{%
\bar{j}},~\ \ dF=0,
\end{equation*}
there exists a horizontal form $A$ such that $F=dA.$

\bigskip

\section{Inhomogeneous Maxwell equations}

As we have seen, in Finsler spaces the 4-potential $A$ and the generalized
Faraday 2-form are defined on the tangent bundle $TM.$

The interaction term of the total action becomes%
\begin{equation*}
S_{int}=-\sum \dfrac{q}{c}\int A=-\sum \dfrac{q}{c}\int A_{i}(x,\dot{x}%
)dx^{i}.
\end{equation*}

In the classical Riemannian case, the above integral is transformed into one
on a domain in the spacetime $M$. In our case, we will transform it into an
integral on a domain in $TM.$ That is, we must write total charge as an
integral (on a domain as specified above):

\begin{equation*}
q=\int \dfrac{\rho (x)}{\sqrt{g_{00}}}\sqrt{G}d^{3}x\wedge d^{4}y.
\end{equation*}

This way, $S_{int}$ will be given by%
\begin{equation*}
-\dfrac{1}{c}\int A_{i}\dfrac{\rho (x)}{\sqrt{g_{00}}}\dfrac{dx^{i}}{dx^{0}}%
\sqrt{G}d^{4}x\wedge d^{4}y
\end{equation*}

With the notation%
\begin{equation}
J^{i}=\dfrac{\rho c}{\sqrt{g_{00}}}\dfrac{dx^{i}}{dx^{0}},
\label{horizontal_J}
\end{equation}%
the integral $\int A_{k}dx^{k}$ is written as%
\begin{equation}
-\dfrac{q}{c}\int A_{k}dx^{k}=-\dfrac{1}{c}\int A_{i}J^{i}d\Omega .
\label{interaction_Lagrangian}
\end{equation}

The above expression is formally similar to the one in the pseudo-Riemannian
case, though, here, the volume element is considered on a certain domain in
the tangent bundle $TM.$

The quantities $J^{i}$ (interpreted as components of the 4-current) thus
define the horizontal component of some vector field%
\begin{equation*}
J=J^{i}\delta _{i}+J^{\bar{\imath}}\partial _{\bar{\imath}}
\end{equation*}%
on $TM.$

\bigskip

The field equations can be obtained by varying with respect to the potential 
$A$ the action%
\begin{equation}
S_{1}=-\dfrac{1}{c}\int A-\dfrac{1}{16\pi c}\int F\ast Fd^{4}x\wedge d^{4}y.
\label{field_action}
\end{equation}

This variation leads to:%
\begin{equation}
\dfrac{1}{\sqrt{G}}\{(F^{ij}\sqrt{G})_{;j}-F^{ij}N_{j\cdot \bar{k}}^{\bar{k}}%
\sqrt{G}\}+\dfrac{1}{\sqrt{G}}(\tilde{F}^{i\bar{j}}\sqrt{G})_{\cdot \bar{j}%
}=-\dfrac{4\pi }{c}J^{i}.  \label{Max_currents}
\end{equation}

With $v=v(x)$ and $N_{~j}^{\bar{\imath}}(x,y)=\overset{(v)}{\gamma }\overset{%
}{_{~jk}^{\bar{\imath}}(x)}y^{k},$ we have%
\begin{equation}
\dfrac{1}{\sqrt{\left\vert g\right\vert }}\{\delta _{j}(F^{ij}\sqrt{%
\left\vert g\right\vert })+(F^{i\bar{j}}\sqrt{\left\vert g\right\vert }%
)_{\cdot \bar{j}}\}=-\dfrac{4\pi }{c}J^{i}.  \label{Max2}
\end{equation}

\bigskip

\textbf{Notes: }1) In the integral above, in order to make sure that the
expression has physical sense, we might need to adjust measurement units so
as to have $[F_{ij}]=[F_{i\bar{j}}].$ This can be done, by considering the
fiber coordinates $y^{\bar{\imath}}$ as having the same measurement units as
the base ones (eventually, by multiplying them by a constant, \cite{animmath}%
).

2) We remark a certain resemblance between the term $(F^{i\bar{j}}\sqrt{G}%
)_{\cdot \bar{j}}$ and the idea of bound current in a material medium.

\bigskip

Equations (\ref{Max2}) gave the idea to formally generalize the \textit{%
inhomogeneous Maxwell equation} as 
\begin{equation}
\delta F=-\dfrac{4\pi }{c}J_{\flat }.
\end{equation}

\bigskip

In local coordinates, this is:%
\begin{eqnarray}
F_{~~|j}^{ij}+F_{~~~\cdot \bar{j}}^{i\bar{j}}+Q^{i} &=&-\dfrac{4\pi }{c}%
J^{i},  \label{inhom_max_local} \\
F_{~~|j}^{\bar{\imath}j}+Q^{\bar{\imath}} &=&-\dfrac{4\pi }{c}J^{\bar{\imath}%
},  \notag
\end{eqnarray}%
where 
\begin{eqnarray}
Q^{i} &=&-F^{ij}P_{~j\bar{k}}^{\bar{k}}+F^{i\bar{j}}\partial _{\bar{j}}(\ln 
\sqrt{\left\vert G\right\vert })  \label{Q} \\
Q^{\bar{\imath}} &=&-\dfrac{1}{2}F^{jk}R_{~jk}^{\bar{\imath}}-F^{j\bar{k}%
}P_{~j\bar{k}}^{\bar{\imath}}-F^{\bar{\imath}j}P_{~j\bar{k}}^{\bar{k}}. 
\notag
\end{eqnarray}

In particular, if $v=v(x)$ and $N_{~j}^{\bar{\imath}}(x,y)=\overset{(v)}{%
\gamma }\overset{}{_{~jk}^{\bar{\imath}}(x)}y^{k},$ this yields%
\begin{eqnarray}
&&F_{~~|j}^{ij}+F_{~~\cdot \bar{j}}^{i\bar{j}}+F^{i\bar{j}}C_{~l\bar{j}%
}^{l}=-\dfrac{4\pi }{c}J^{i}  \label{inhomogeneous_max} \\
&&F_{~~|j}^{\bar{\imath}j}-\dfrac{1}{2}F^{jk}R_{~jk}^{\bar{\imath}}=-\dfrac{%
4\pi }{c}J^{\bar{\imath}},  \notag
\end{eqnarray}

The first set of equations is nothing but (\ref{inhomogeneous_max}) obtained
by variational methods, while the second one is new. We notice the
appearance of the quantities $J^{\bar{\imath}}$ (due to both the Finslerian
character of the space and the nonholonomy of the frame we used) which are
"coupled" on $TM$ to the usual components of the 4-current $J^{i}.$

In the following, we will see that $J^{\bar{\imath}}$ play an important role
in the continuity equation and in the Finslerian analogue of energy-momentum
conservation law.

\section{Continuity equation and gauge invariance}

Above, we have seen that 
\begin{equation}
-\dfrac{4\pi }{c}J_{\flat }=\delta F.  \label{current_TM}
\end{equation}%
There immediately follows: $-\dfrac{4\pi }{c}\delta J_{\flat }=\delta \delta
F=0$\footnote{%
We have used the identity $\delta \delta \omega =(-1)^{2p}(\ast ^{-1}d\ast
)(\ast ^{-1}d\ast )\omega =\ast ^{-1}dd\omega =0.$}, which is, $\ div(J)=0.$
In other words:

\begin{proposition}
There holds the generalized continuity equation:%
\begin{equation}
div(J)=0.  \label{continuity_eq}
\end{equation}
\end{proposition}

We have seen above that the electromagnetic tensor $F$ is invariant under
transformations $A(x,y)~\mapsto \tilde{A}(x,y):=A(x,y)+d\lambda (x)$ of the
4-potential. It means that, in the general action (\ref{general_Lagrangian}%
), the first term $S_{p}$ and the third one $S_{f}$ will also be invariant.

The continuity equation (\ref{continuity_eq}) insures that, with respect to
the above transformations, $\tilde{S}_{int}=-\int \tilde{A}_{i}J^{i}\sqrt{%
\left\vert G\right\vert }d^{4}x\wedge d^{4}y$ equals $S_{int}$ plus a
boundary term.

Indeed\textit{, }we have (omitting the minus sign in front of the integral):%
\begin{equation*}
\int \tilde{A}_{i}J^{i}d\Omega =\int (A_{i}+\dfrac{\partial \lambda }{%
\partial x^{i}}J^{i})\sqrt{\left\vert G\right\vert }d^{4}x\wedge d^{4}y.
\end{equation*}%
Since $\lambda $ depeds only on $x$, we can write $\dfrac{\partial \lambda }{%
\partial x^{i}}=\lambda _{;i},$ hence the term to be added to $\int
A_{i}J^{i}\sqrt{\left\vert G\right\vert }d^{4}x\wedge d^{4}y$ is 
\begin{equation*}
\int \dfrac{\partial \lambda }{\partial x^{i}}J^{i}\sqrt{\left\vert
G\right\vert }d^{4}x\wedge d^{4}y=\int (\lambda J^{i}\sqrt{\left\vert
G\right\vert })_{;i}~d^{4}x\wedge d^{4}y-\int \lambda (J^{i}\sqrt{\left\vert
G\right\vert })_{;i}~d^{4}x\wedge d^{4}y
\end{equation*}

This term can be written as $\int div(\lambda J^{H})d\Omega -\int \lambda
div(J^{H})d\Omega .$ According to the continuity equation and taking into
account that $\lambda $ does not depend on $y$, we can write it as $\int
div(\lambda J^{H})d\Omega +$ $\int div(\lambda J^{V})d\Omega =\int
div(\lambda J)d\Omega ,$ i.e., it can be written as a boundary term. When
performing variations of the action (and assuming, as in the classical case,
that variations vanish on the boundary), these terms will cancel out.

\bigskip

In conclusion, transformations $A(x,y)~\mapsto A(x,y)+d\lambda (x)$ of the
4-potential do not affect the action (\ref{general_Lagrangian}).

\textbf{Remark. }If $A_{i}=A_{i}(x),$ then from (\ref{inhomogeneous_max}),
it follows $J^{\bar{\imath}}=0.$

\bigskip

\section{Equations of motion}

Let us consider momentarily the case of a single particle. The equations of
motion are obtained by varying the trajectory $x=x(t)$ in the first two
terms of (\ref{general_Lagrangian}), which are in our case written in the
form of a single integral along the considered curve:%
\begin{equation}
S_{2}=-\int {\Large (}mc\sqrt{g_{ij}(x,\dot{x})\dot{x}^{i}\dot{x}^{j}}+%
\dfrac{q}{c}A_{k}(x,\dot{x})\dot{x}^{k}{\Large )}dt.
\label{Lorentz_Lagrangian}
\end{equation}

The 0-homogeneity of $A$ insures that the action $S_{2}$ is invariant under
eventual changes of parameter $t\mapsto t^{\prime }$ of the curve.

\bigskip

A further restriction can be imposed on the $y$-dependence of $A$ in order
to make all the approach more elegant and provide a simple relation of $A$
with the canonical 4-momentum and the equations of motion of charged
particles.

Once the independence of the integral on the parametrization was
established, we are free to choose the parameter along the considered
curves. Traditionally, when deducing the equations of motion, curves are
parametrized by the arclength $s.$ In this case, the action $S_{2}$ in (\ref%
{Lorentz_Lagrangian}) is equivalent to the one provided by the Lagrangian%
\begin{equation}
L=\dfrac{1}{2}mcg_{ij}(x,y)y^{i}y^{j}+\dfrac{q}{c}A_{k}(x,y)y^{k},~\ \ y=%
\dfrac{dx}{ds},  \label{Lorentz_L_improved}
\end{equation}%
which is more comfortable in view of Legendre duality and Hamiltonian
formalism.

The canonical momentum of $L$ is given by 
\begin{equation*}
p_{i}=\dfrac{\partial L}{\partial y^{i}}=mcy_{i}+\dfrac{q}{c}(A_{k\cdot
i}y^{k}+A_{i}).
\end{equation*}

\bigskip

In \textit{isotropic }(pseudo-Riemannian) spaces, if we assume that $A=A(x),$
then there exists only one potential providing a given interaction
Lagrangian $L_{int}=A_{i}(x)y^{i}$. But in Finsler spaces, where $%
A_{i}=A_{i}(x,y),$ a given Lagrangian $L_{int}=A_{i}(x,y)y^{i}$ can be given
by infinitely many functions $A_{i}=A_{i}(x,y).$ Thus, to a Lagrangian $%
L_{int},$ it corresponds a whole equivalence class of potentials $A.$
Comparing to (\ref{canonical_momentum}), it appears as convenient to choose
from each class the representative for which

\begin{equation}
A_{k\cdot i}y^{k}=0.  \label{representative_A}
\end{equation}

We will call this condition upon $A,$ the \textit{gradient gauge. }In the
gradient gauge, 
\begin{equation*}
A_{i}=\dfrac{\partial (A_{k}y^{k})}{\partial y^{i}}.
\end{equation*}

\textbf{Remark. }0-homogeneity of $A$ insures that we also have $A_{i\cdot
k}y^{k}=0.$

\bigskip

In the gradient gauge, the canonical 4-momentum is given by%
\begin{equation*}
p_{i}=\dfrac{\partial L}{\partial y^{i}}=mcy_{i}+\dfrac{q}{c}A_{i},
\end{equation*}%
in other words, the \textit{Liouville (canonical) 1-form} $\theta =\dfrac{%
\partial L}{\partial y^{i}}dx^{i}$ attached to $L$ is given by $\theta
=(mcy_{i}+\dfrac{q}{c}A_{i})dx^{i}$ and the \textit{Poincar\'{e} 2-form} $%
\omega =d\theta ,$ by 
\begin{equation*}
\omega =\dfrac{1}{2}(A_{j|i}-A_{i|j})dx^{i}\wedge dx^{j}-(mcg_{ij}+\dfrac{q}{%
c}A_{i\cdot j})dx^{i}\wedge \delta y^{\bar{j}}
\end{equation*}

\bigskip

In the following, we assume that the tangent bundle $TM$ is endowed with an
(arbitrary) Ehresmann connection $N$ and the corresponding linear connection 
$D\Gamma (N),$ (\ref{Chern}).

Variation of (\ref{Lorentz_L_improved}) provides the Euler-Lagrange
equations:%
\begin{equation}
mc\dfrac{Dy^{i}}{ds}=\dfrac{q}{c}F_{~j}^{i}y^{j}+\dfrac{q}{c}F_{~\bar{j}}^{i}%
\dfrac{\delta y^{\bar{j}}}{ds},~\ \ y^{i}=\dfrac{dx^{i}}{ds},
\label{Lorentz}
\end{equation}%
where $\dfrac{Dy^{i}}{ds}=\dfrac{dy^{i}}{ds}+L_{~jk}^{i}y^{j}y^{k}$ and we
assumed $A_{k\cdot i}y^{k}=0.$

\bigskip

The first term in the right hand side above is similar to the usual one in
pseudo-Riemannian spaces, while the second one $\dfrac{q}{c}F_{~\bar{j}}^{i}%
\dfrac{\delta y^{\bar{j}}}{ds}$ is new and appears due to the dependence of $%
A$ on the variable $y.$

\bigskip

\textbf{Remark. }Both the "traditional" Lorentz force term (given by $F^{i}=%
\dfrac{q}{c}F_{~h}^{i}y^{h}$) and the correction given by $\tilde{F}^{i}=%
\dfrac{q}{c}F_{~\bar{j}}^{i}\dfrac{\delta y^{\bar{j}}}{ds}$ are orthogonal
to the velocity 4-vector $y=\dot{x}:$%
\begin{equation}
g_{ij}F^{i}y^{j}=0,~~g_{ij}\tilde{F}^{i}y^{j}=0.  \label{ortho}
\end{equation}

\textbf{Remark. }In equations (\ref{Lorentz}), we can use any Ehresmann
connection $N,$ their form does not depend on $N$.

\section{Stress-energy-momentum tensor}

\subsection{In flat pseudo-Finsler spaces}

Let us consider the vector space $M=\mathbb{R}^{4}$ endowed with a flat
pseudo-Finsler metric%
\begin{equation*}
g_{ij}=g_{ij}(y).
\end{equation*}%
Assuming that coordinate transformations are linear (as traditionally done
in special relativity), we can choose the trivial Ehresmann connection $%
N_{~j}^{\bar{\imath}}=0,$ hence the adapted frame on $TM$ is the natural one 
$(\delta _{i}=\dfrac{\partial }{\partial x^{i}},$ $\partial _{\bar{\imath}}=%
\dfrac{\partial }{\partial y^{\bar{\imath}}})$ and its dual is $%
(dx^{i},\delta y^{\bar{\imath}}=dy^{\bar{\imath}}).$

Spacetime translations $\bar{x}^{i}=x^{i}+\varepsilon ^{i},$ $i=\overline{0,3%
}$ induce the following transformation on $TM:$%
\begin{equation}
\bar{x}^{i}=x^{i}+\varepsilon ^{i},~\ \bar{y}^{i}=y^{i}.
\label{translations}
\end{equation}

By \textit{generalized energy-momentum tensor on }$TM$, we understand the
Noether current given by the invariance to spacetime translations
(accordingly, to transformations (\ref{translations})) of the action 
\begin{equation}
S_{F}=-\int \dfrac{1}{16\pi c}F\ast Fd^{4}x\wedge d^{4}y,
\label{field_action1}
\end{equation}%
symmetrized by adding a divergence term.

\bigskip

As pointed out in Section 2, invariance with respect to transformations (\ref%
{translations}) of the above means the absence of explicit dependence on the
base coordinates $x^{i}$ of the Lagrangian.

In order to find the form of the energy-momentum tensor of the
electromagnetic field in a flat pseudo-Finslerian space, let us first see
how Noether theorem is read in these spaces.

\bigskip

For an action 
\begin{equation}
S=\dfrac{1}{c}\int \Lambda (q_{(l)},\dfrac{\partial q_{(l)}}{\partial x^{i}},%
\dfrac{\partial q_{(l)}}{\partial y^{i}})d\Omega ,  \label{Finslerian_action}
\end{equation}%
where $\Lambda =L\sqrt{\left\vert G\right\vert }$ is a Lagrangian density on 
$TM$ and $q_{(l)}=q_{(l)}(x,y)$ are the field variables, the Euler-Lagrange
equations are:%
\begin{equation}
\dfrac{\partial }{\partial x^{i}}(\dfrac{\partial \Lambda }{\partial
q_{(l),i}})+\dfrac{\partial }{\partial y^{i}}(\dfrac{\partial \Lambda }{%
\partial q_{(l)\cdot i}})-\dfrac{\partial \Lambda }{\partial q_{(l)}}=0.
\label{Euler-Lagrange}
\end{equation}

The absence of explicit dependence on $x^{i}$ of $\Lambda $ means%
\begin{equation*}
\dfrac{\partial \Lambda }{\partial x^{i}}=\dfrac{\partial \Lambda }{\partial
q_{(l)}}\dfrac{\partial q_{(l)}}{\partial x^{i}}+\dfrac{\partial \Lambda }{%
\partial q_{(l),k}}q_{(l),ki}+\dfrac{\partial \Lambda }{\partial q_{(l)\cdot
k}}q_{(l)\cdot k~,i}.
\end{equation*}%
(where we understood also summation over $l$). Substituting $\dfrac{\partial
\Lambda }{\partial q_{(l)}}$ from (\ref{Euler-Lagrange}) and grouping terms,
we get%
\begin{equation*}
\dfrac{\partial }{\partial x^{k}}\left( q_{(l),i}\dfrac{\partial \Lambda }{%
\partial q_{(l),k}}-\delta _{i}^{k}\Lambda \right) +\dfrac{\partial }{%
\partial y^{\bar{k}}}\left( q_{(l),i}\dfrac{\partial \Lambda }{\partial
q_{(l)\cdot \bar{k}}}\right) =0.
\end{equation*}

Thus, the invariance of an action on $TM$ under translations on the (flat)
base space $M$ leads to the appearance of a quantity consisting of \textit{%
two} \textit{blocks}, namely, by symmetrizing (adding divergence terms to)
the quantities%
\begin{equation}
\tilde{T}_{~i}^{k}=\dfrac{1}{\sqrt{\left\vert G\right\vert }}q_{(l),i}\dfrac{%
\partial \Lambda }{\partial q_{(l),k}}-\delta _{i}^{k}\Lambda ,~\ \ \ \ 
\tilde{T}_{~i}^{\bar{k}}=\dfrac{1}{\sqrt{\left\vert G\right\vert }}q_{(l),i}%
\dfrac{\partial \Lambda }{\partial q_{(l)\cdot \bar{k}}}
\label{general_em_tensor}
\end{equation}

With these notations,%
\begin{equation*}
\dfrac{\partial }{\partial x^{k}}\left( \tilde{T}_{~i}^{k}\sqrt{\left\vert
G\right\vert }\right) +\dfrac{\partial }{\partial y^{\bar{k}}}\left( \tilde{T%
}_{~i}^{\bar{k}}\sqrt{\left\vert G\right\vert }\right) =0.
\end{equation*}

\bigskip

\textbf{Case 1 }$(J=0).$ In order to "guess" the form of the generalized
energy-momentum tensor for the electromagnetic field, it is advantageous to
assume for the beginning that $J=0$ and apply the above construction to the
Lagrangian density:%
\begin{equation*}
\Lambda =-\dfrac{1}{16\pi }F\ast F=-\dfrac{1}{16\pi }F_{BC}F^{BC}\sqrt{%
\left\vert G\right\vert }.
\end{equation*}

\bigskip

We get $\dfrac{\partial \Lambda }{\partial A_{k,l}}=-\dfrac{1}{4\pi }F^{lk}%
\sqrt{\left\vert G\right\vert },~\ \ \dfrac{\partial \Lambda }{\partial
A_{k\cdot \bar{l}}}=-\dfrac{1}{4\pi }F^{\bar{l}k}\sqrt{\left\vert
G\right\vert }\ $and%
\begin{equation*}
\tilde{T}_{~i}^{l}=\dfrac{1}{4\pi }(-F^{lk}A_{k,i}+\dfrac{1}{4}\delta
_{i}^{l}F_{BC}F^{BC}),\tilde{T}_{~i}^{\bar{l}}=-\dfrac{1}{4\pi }F^{\bar{l}%
k}A_{k,i}.
\end{equation*}

The obtained Noether current can be symmetrized by adding divergence terms.
It can be easily seen that, for $J=0,$ the term $\dfrac{1}{4\pi }%
(F^{lk}A_{i,k}+F^{l\bar{k}}A_{i\cdot \bar{k}})$ can be expressed as a
divergence term.

By adding this term to $\tilde{T}_{~i}^{l}$, we get $T_{~i}^{l}=-\dfrac{1}{%
4\pi }(F^{lk}F_{ik}+F^{l\bar{k}}F_{i\bar{k}}-\delta _{i}^{l}F_{BC}F^{BC})$
or, equivalently,%
\begin{equation}
T_{~i}^{l}=\dfrac{1}{4\pi }(-F^{lB}F_{iB}+\dfrac{1}{4}\delta
_{i}^{l}F_{BC}F^{BC}).  \label{horizontal_sem_tensor}
\end{equation}

Similarly, $\dfrac{1}{4\pi }F^{\bar{l}k}A_{i,k}$ can be expressed as a
divergence term; by adding it to $\tilde{T}_{~i}^{\bar{l}}=-\dfrac{1}{4\pi }%
F^{\bar{l}k}A_{k,i}$ , we obtain%
\begin{equation}
T_{~i}^{\bar{l}}=-\dfrac{1}{4\pi }F^{\bar{l}k}F_{ik}.
\label{mixed_sem_tensor}
\end{equation}

Thus, in the case $J=0$, we get $div(T)=0,$ i.e., 
\begin{equation*}
\dfrac{\partial }{\partial x^{k}}\left( T_{~i}^{k}\sqrt{\left\vert
G\right\vert }\right) +\dfrac{\partial }{\partial y^{\bar{k}}}\left( T_{~i}^{%
\bar{k}}\sqrt{\left\vert G\right\vert }\right) =0.
\end{equation*}

\bigskip

This suggests the following

\begin{definition}
The generalized energy-momentum tensor in the flat Finsler space $(\mathbb{R}%
^{4},\mathcal{F}(y))$ is the symmetric tensor%
\begin{equation}
T=T_{ij}dx^{i}\otimes dx^{j}+T_{i\bar{j}}dx^{i}\otimes dy^{\bar{j}}
\label{sem_tensor}
\end{equation}%
with local components given by (\ref{horizontal_sem_tensor}) and (\ref%
{mixed_sem_tensor}).
\end{definition}

The horizontal component $T_{ij}dx^{i}\otimes dx^{j}$ is the the usual
energy-momentum tensor (plus some correction due to anisotropy), while the
mixed one $T_{i\bar{j}}dx^{i}\otimes dy^{\bar{j}}$ is new. As we have seen
above, these new components play a role in the analogue of the conservation
law.

\bigskip

\textbf{Case 2 }$(J\not=0).$ Let now the $TM$-current $J$ be arbitrary. By
using Maxwell equations (with $N=0$), we get:%
\begin{equation}
\dfrac{1}{\sqrt{\left\vert G\right\vert }}[\dfrac{\partial }{\partial x^{j}}%
(T_{~i}^{j}\sqrt{\left\vert G\right\vert })+\dfrac{\partial }{\partial y^{%
\bar{j}}}(T_{~i}^{\bar{j}}\sqrt{\left\vert G\right\vert })]=-\dfrac{1}{c}%
(F_{ij}J^{j}+F_{i\bar{j}}J^{\bar{j}}).  \label{sem_conservation}
\end{equation}%
In brief, 
\begin{equation}
div(T)=-\dfrac{1}{c}i_{J}F.  \label{sem_conservation_brief}
\end{equation}

\subsection{In general Finsler spaces}

In general (pseudo-)Finsler spaces, we will still define the generalized
energy-momentum tensor for the electromagnetic field as above:%
\begin{eqnarray}
T &=&T_{ij}dx^{i}\otimes dx^{j}+T_{i\bar{j}}dx^{i}\otimes \delta y^{\bar{j}%
},~\ \   \label{sem_tensor_general} \\
T_{~iA} &=&\dfrac{1}{4\pi }(-F_{A}^{~~B}F_{iB}+\dfrac{1}{4}%
g_{iA}F_{BC}F^{BC}),  \notag
\end{eqnarray}%
where $g_{i\bar{j}}=0$ and indices $A,B,C$ take all values corresponding to
both horizontal and vertical components.

\bigskip

\textbf{Remark. }The horizontal components $T_{ij}$ of the generalized
energy-momentum tensor can be obtained by varying the action $S_{F}$ with
respect to the spacetime metric $g$ (i.e., with respect to the \textit{%
horizontal} part of the metric $(G_{AB})$):%
\begin{equation*}
\delta _{g}S_{F}=\dfrac{1}{2c}\int T_{ik}\delta g^{ik}d\Omega ,
\end{equation*}%
while the mixed components $T_{i\bar{j}}$ are obtained by varying
(independently) $S_{F}$ with respect to the Ehresmann connection $N:$%
\begin{equation*}
\delta _{N}S_{F}=\dfrac{1}{c}\int T_{~\bar{\imath}}^{j}\delta N_{~j}^{\bar{%
\imath}}d\Omega .
\end{equation*}

\bigskip

In curved pseudo-Finsler spaces case, the adapted frame $(\delta
_{i},\partial _{\bar{\imath}})$ is generally nonholonomic, hence any linear
connection $D\Gamma (N)$ which preserves the distributions generated by $N$
has generally nonvanishing torsion (at least, $\delta y^{\bar{\imath}%
}(T(\delta _{k},\delta _{j}))=R_{~jk}^{\bar{\imath}}\not=0$). In this case,
the covariant divergence of the energy-momentum tensor is not simply equal
to $-\dfrac{1}{c}i_{J}F$, but has a more complicated expression, involving
the torsion tensor. The situation formally resembles to the one in
Riemann-Cartan geometry, \cite{watanabe}.

In order to find the relation between $\dfrac{1}{c}i_{J}F$ and the
generalized energy-momentum tensor, it appears as most comfortable to
express $-\dfrac{1}{c}i_{J}F=-\dfrac{1}{c}(F_{ij}J^{j}+F_{i\bar{j}}J^{\bar{j}%
})$ in terms of covariant derivatives (\ref{Chern}).

\bigskip

Let us assume for simplicity that $v=v(x)$ and $N$ is given by (\ref%
{special_connection}). Then, 
\begin{equation}
Q^{i}=F^{i\bar{j}}C_{h\bar{j}}^{h},~\ \ Q^{\bar{\imath}}=-\dfrac{1}{2}%
F^{jk}R_{~jk}^{\bar{\imath}}.
\end{equation}

Taking into account the Maxwell equations, we get%
\begin{equation}
-\dfrac{1}{c}(F_{ij}J^{j}+F_{i\bar{j}}J^{\bar{j}})=T_{~i|j}^{j}+T_{~i\cdot 
\bar{j}}^{\bar{j}}+T_{~i}^{\bar{j}}C_{~h\bar{j}}^{h}+T_{~\bar{k}%
}^{j}R_{~ij}^{\bar{k}}{\Large .}
\end{equation}

\section{Conclusion}

For a 4-dimensional pseudo-Finsler space $(M,\mathcal{F})$, we have
constructed a notion of electromagnetic tensor, based exclusively on
variational calculus and exterior derivative, \cite{animmath}.

The 4-potential is defined as a horizontal 1-form $A=A_{i}(x,y)dx^{i}\ $on
the tangent bundle $TM,$ having its components $A_{i}$ homogeneous of degree
0 in $y.$

In terms of this 4-potential, the generalized electromagnetic tensor is the
2-form $F=dA.$ The Maxwell equations on $TM$ are then written as:%
\begin{equation*}
dF=0,~\ \ \delta F=-\dfrac{4\pi }{c}J_{\flat }.
\end{equation*}

The $TM$-current $J=J^{i}\delta _{i}+J^{\bar{\imath}}\partial _{\bar{\imath}%
} $ is a vector field on $TM$ satisfying identically $div~J=0.$ Its
horizontal component $J^{i}\delta _{i}$ provides the usual notion of
4-current (plus a correction term due to the $y$-dependence of $A$), while
the vertical one $J^{\bar{\imath}}\partial _{\bar{\imath}}$ is due to the
anisotropy of the 4-potential and to the nonholonomy of the frame.

Further, for flat pseudo-Finsler spaces $(M,\mathcal{F}(y))$, the
generalized energy-momentum tensor is defined as the symmetrized Noether
current corresponding to invariance to spacetime translations of the field
Lagrangian. We obtained%
\begin{eqnarray}
T &=&T_{ij}dx^{i}\otimes dx^{j}+T_{i\bar{j}}dx^{i}\otimes dy^{\bar{j}}, \\
T_{~iA} &=&\dfrac{1}{4\pi }(-F_{A}^{~~B}F_{iB}+\dfrac{1}{4}\delta
_{A}^{l}F_{BC}F^{BC}),
\end{eqnarray}%
(where $\delta _{j}^{l}$ is the Kronecker delta and $\delta _{\bar{j}}^{l}=0$
and $A,B,C$ take all values corresponding to both horizontal and vertical
components). The generalized energy-momentum tensor satisfies the
conservation law%
\begin{equation*}
div(T)=-\dfrac{1}{c}(F_{ij}J^{j}+F_{i\bar{j}}J^{\bar{j}}).
\end{equation*}

In curved Finsler spaces, the same expressions can be obtained by varying by
varying the action $S_{F}=-\int \dfrac{1}{16\pi }F\ast Fd\Omega $ for the
field with respect to the metric $g_{ij}(x,y)$ (thus getting $T_{ij}$) and
with respect to the Ehresmann connection $N$ (which provides the components, 
$T_{i\bar{j}}).$

The above considerations hold true in more general spaces such as Lagrange
or generalized Lagrange spaces, with the only mention that in these spaces,
the equations of motion of charged particles become more complicated (and
the 0-homogeneity assumption on $A$ is dropped).

\textbf{Remark. }Prior to this model, to our knowledge, there existed only
one geometric model for electromagnetism in Finsler spaces, belonging to R.
Miron and collaborators, \cite{Lagrange}, \cite{Miron-Rad}, \cite%
{Miron-Buchner}, \cite{Ingarden}, in which it is defined, by means of
deflection tensors of linear connections on $TM$, a notion of
electromagnetic tensor on $TM$ (with horizontal \textit{hh-} and vertical 
\textit{vv-} components) and it is provided a generalization of Maxwell
equations. Several of the the advantages of our approach are: obtaining by
variational methods the $TM$-versions of: Maxwell equations, equations of
motion (and Lorentz force, respectively), energy-momentum tensor; an easier
interpretation of the new (mixed $hv$-) component of the electromagnetic
2-form (as appearing, for instance, in the equations of motion); obtaining
an analogue of the usual continuity equation as an identity, and also, the
possibility of a compact writing using exterior derivatives.


\bigskip

\begin{thebibliography}{99}
\bibitem{Asanov} Asanov, G.S. , \textit{Finsler Geometry, Relativity and
Gauge Theories} , Reidel, Dordrecht, 1985.

\bibitem{Rahula} Atanasiu, Gh., Balan, V., Br\^{\i}nzei, N., Rahula, \textit{%
Differential-geometric structures -- tangent bundles, connections in
bundles, exponential law and jet spaces, }"Librokom", Moscow, 2009 (\textit{%
in Russian).}

\bibitem{Bertschinger} Bertschinger, E., \textit{Symmetry transformations,
the Einstein-Hilbert action and gauge invariance, }MIT, 2002.

\bibitem{Ba-St1} Balan, V., Stavrinos, P.C., \textit{Weak gravitational
fields in generalized metric spaces}, Proc. of The Int. Conf. of Geometry
and Its Applications, Thessaloniki, Greece 1999, BSG Proc. 6, Geometry
Balkan Press, Bucharest, 2002, 27-37.

\bibitem{Ba-St2} Balan, V., Stavrinos, P.C., Trencevski, K., \textit{Weak
gravitational models based on Beil metrics}, Proc. of the Conference of
Applied Differential Geometry -General Relativity,Workshop "Applied
Differential Geometry, Lie Algebras and General Relativity", August 27 -
September 2, 2000, Thessaloniki, Greece.

\bibitem{Ba-St3} Balan, V., Stavrinos, P.C., \textit{Weak linearized
gravitational models based on Finslerian }$(\alpha ,\beta )$\textit{-metrics}%
, Proc. of The Conference of Applied Differential Geometry - General
Relativity - June 26 - July 1, 2001, Thessaloniki, Greece.

\bibitem{Ba-St4} Balan, V., Stavrinos, P.C., \textit{Finslerian (}$\alpha
,\beta $\textit{)-metrics in weak gravitational models}, in "Finsler and
Lagrange Geometries", Proc. of The Conference held on Aug. 26-31 2001 in
Iasi, Romania, Eds: M.Anastasiei and P.L.Antonelli, Kluwer Acad. Publishers
2003, 259-268.

\bibitem{Bogoslovski1} Bogoslovsky G.Yu., Goenner H. F., \textit{On the
possibility of phase transitions in the geometric structure of space-time},
Phys. Lett. A, 1998, V. 244, 222--228.

\bibitem{Bogoslovski 2} Bogoslovsky G.Yu., \textit{A viable model of locally
anisotropic space-time and the Finslerian generalization of the relativity
theory}, Fortschr. Phys., 1994, V. 42, N 2, 143--193.

\bibitem{Bogoslovski3} Bogoslovsky G.Yu., Goenner H. F., \textit{Concerning
the generalized Lorentz symmetry and the generalization of the Dirac equation%
}, Phys. Lett. A, 2004, V. 323, 40-47.

\bibitem{Cairo08} N.Brinzei (Voicu), S.Siparov, \textit{Equations of
electromagnetism in some special anisotropic spaces,} arXiv:0812.1513v1
[gr-qc], 08 Dec. 2008.

\bibitem{Shen} Bao, D., Chern, S.S., Shen, Z, \textit{An Introduction to
Riemann-Finsler Ge\-o\-me\-try} (Graduate Texts in Mathematics; 200),
Springer Verlag, 2000.

\bibitem{Rham1} Fortini, P., Montanari, E., Ortolan, A., Schafer, G., 
\textit{Gravitational Wave Interaction with Normal and Superconducting
Circuits}, arXiv:gr-qc/9808080v1, 1998.

\bibitem{garas'ko} Garas'ko, G.I., \textit{Fundamentals of Finsler Geometry
for Physicists} (in Russian),Tet-ru Eds., Moscow 2009

\bibitem{Lagrange} Miron, R., Anastasiei, M., \textit{The Ge\-o\-me\-try of
Lagrange\- Spaces: Theory and Applications}, Kluwer Acad. Publ. FTPH no. 59,
(1994).

\bibitem{Landau} Landau, L.D., Lifschiz, E.M., \textit{Field Theory}, 8th
ed., Fizmatlit, Moscow, 2006\textit{.}

\bibitem{Miron-Buchner} Miron, R., Rosca, R., Anastasiei, M., Buchner, K., 
\textit{New aspects in Lagrangian relativity}, Found. of Phys. Lett. 2, 5
(1992), 141-171.

\bibitem{Miron-Rad} Miron, R., Radivoiovici-Tatoiu, M., \textit{A Lagrangian
theory of electromagnetism}, Seminarul de Mecanica, Timisoara, 1988, pp.
1-55.

\bibitem{Ingarden} Miron, R., \textit{The geometry of Ingarden spaces, }Rep.
on Math. Phys., 54(2), 2004, pp. 131-147

\bibitem{Pavlov} Pavlov, D.G. (ed.), \textit{Space-Time Structure. Collected
papers}, ed. Tetru, Moscow, 2006.

\bibitem{RBS} Raigorodski, L.D., Stavrinos, P.C., Balan, V., \textit{%
Introduction to the Physical Principles of Differential Geometry, }Univ. of
Athens, 1999.

\bibitem{Rutz} Rutz, S., \textit{A Finsler generalisation of Einstein's
vacuum field equations}, General Relativity and Gravitation, Vol 25 (11),
1993, pp.1139-1158.

\bibitem{agd} Siparov, S.: \textit{On the interpretation of the classical
GRT\ tests and cosmological constant in anisotropic geometrodynamics, }%
arXiv: 0910.3408, 2009.

\bibitem{Shen1} Shen, Z.: \textit{Lectures on Finsler Geometry, }World
Scientific, 2001.

\bibitem{Szilasi} Szilasi, J., \textit{Calculus along the tangent bundle
projection and projective metrizability, }Diff. Geom. and Appl., Proc.
Conf., in Honour of Leonhard Euler, Olomouc, August 2007.

\bibitem{Balan-Udriste} C.Udriste, V.Balan, \textit{Differential operators
and convexity on vector bundles, endowed with (h; v)-metrics}, An. st. Univ.
"AL.I. Cuza", Sect I, Vol.43, no.1 / 1997, p. 37-50.

\bibitem{vacaru} Vacaru, S., Stavrinos, P., Gaburov, E. and Gonta, D. 
\textit{Clifford and RiemannFinsler Structures in Geometric Mechanics and
Gravity,} Geometry Balkan Press, Bucharest, 2006.

\bibitem{vacaru2} Vacaru, S., \textit{Einstein Gravity, Lagrange--Finsler
Geometry, and Nonsymmetric Metrics,} SIGMA 4 (2008), 071.

\bibitem{hilbert} Voicu, N: \textit{New considerations on Hilbert action and
Einstein equations in anisotropic spaces}, arXiv:0911.5034v1 [gr-qc], 2009.

\bibitem{animmath} Voicu, N., Siparov, S., \textit{A new approach to
electromagnetism in anisotropic spaces}, BSG Proc. 17, 2010, pp. 250-260.

\bibitem{Voicu1} Voicu, N., \textit{On electromagnetism and energy-momentum
tensor of the electromagnetic fieldin spaces with pseudo-Finsler geometry},
preprint.

\bibitem{watanabe} Watanabe, T, Hayashi, M., \textit{General relativity with
torsion, a}rXiv: gr -qc/0409029.

\bibitem{Li} Xin Li, Zhe Chang, \textit{Toward a Gravitation Theory in
Berwald--Finsler Space, }arXiv:0711.1934v1 [gr-qc], 2007.

\bibitem{Zhong} Zhong Chunping, Zhong Tongde, \textit{Horizontal Laplace
operator in real Finsler vector bundles, }Acta Math. Sc.
2008,28B(1):128--140,
\end{thebibliography}
\end{document}